\documentclass[prb,twocolumn,superscriptaddress,showpacs]{revtex4}
\usepackage{graphicx}%
\usepackage{dcolumn}
\usepackage{bm}
\begin{document}

\title{First-principles calculations for the
       adsorption of water molecules on the Cu(100) surface}

\author{Sanwu~Wang}
\altaffiliation{Author to whom correspondence should be addressed. 
       Electronic mails: sanwu.wang@vanderbilt.edu, swang@csit.fsu.edu} 
\affiliation{Department of Physics and Astronomy,
       Vanderbilt University, Nashville, Tennessee 37235, USA.}      
\author{Yanzhao Cao}
\affiliation{School of Computational Science and Information Technology,
       Florida State University, Tallahassee, Florida 32306-4120, USA.}
\affiliation{Department of Mathematics, Florida Agricultural and
       Mechanical University, Tallahassee, Florida 32307, USA.}
\author{P.~A.~Rikvold}
\affiliation{School of Computational Science and Information Technology,
       Florida State University, Tallahassee, Florida 32306-4120, USA.} 
\affiliation{Center for Materials Research and Technology and Department
       of Physics, Florida State University, Tallahassee, Florida
       32306-4350, USA}

\begin{abstract}

First-principles density-functional theory and supercell models are
employed to calculate the adsorption of water molecules on the Cu(100)
surface. In agreement with the experimental observations, the calculations
show that a H$_2$O molecule prefers to bond at a one-fold on-top ($T_1$)
surface site with a tilted geometry. At low temperatures, rotational
diffusion of the molecular axis of the water molecules around the surface
normal is predicted to occur at much higher rates than lateral diffusion
of the molecules. In addition, the calculated binding energy of an
adsorbed water molecule on the surfaces is significantly smaller than the
water sublimation energy, indicating a tendency for the formation of water
clusters on the Cu(100) surface.

\end{abstract}


\maketitle

\section{Introduction}

The adsorption of water on metal surfaces has been extensively studied for
many years due to its technological importance and wide academic
interest.\cite{Thiel,Whitten,Henderson} In recent years, there have even
been some new experiments on water enhanced surface reactions of
environmental importance.\cite{Lu1,Grassian} One of the central issues is
the microscopic understanding of the bonding and orientation
characteristics of water molecules on the surfaces. While experimental and
theoretical investigations have provided unified information about the
bonding structure for most water-adsorbed metal surfaces, theoretical
results are not always in agreement with the experimental observations.
One example is the water-adsorbed Cu(100) surface.%
\cite{Anderson,Spitzer,Nyberg,Sexton,Brosseau,Ribarsky,Kuznetsov,Ignaczak}

Copper is an important material for structural, electrical, and
electronics applications. Corrosion of copper, which is a source of
significant economic losses and occurs both in aqueous environments%
\cite{Stic89,Kear04} and atmospheric air,\cite{Lee97,Aast00} depends
intimately on water adsorption. Adsorption of water on copper is therefore
important, both from technological and basic scientific points of view.
The experimental investigations for the adsorption of water on Cu(100)
have been done  with various techniques including ultraviolet
photoemission spectroscopy (UPS), Auger electron spectroscopy (AES),
electron-energy-loss spectroscopy (EELS), and low-energy electron
diffraction (LEED).\cite{Anderson,Spitzer,Nyberg,Brosseau} Water has been
found to adsorb molecularly on the surface at low temperatures ($<150$~K)
and to tend to form clusters. Temperature programmed desorption (TPD)
measurements have shown that the desorption of water from the surface
occurs at a temperature of $\sim$160-170~K.\cite{Sexton,Brosseau} The EELS
data of Anderson {\it et~al.} suggested that the water molecule is bonded
at the one-fold on-top ($T_1$) site with its oxygen end towards the
surface and its axis tilted away from the surface normal (at an estimated
angle of about $60^\circ$).\cite{Anderson,Nyberg}

Several groups have performed theoretical calculations for the
water-adsorbed Cu (100) surface using cluster models.%
\cite{Ribarsky,Kuznetsov,Ignaczak} While these investigations have
provided useful insight about the interaction between water and the
surface, some of the key results are not in agreement with the
experimental data. For example, while {\it ab initio} density-functional
cluster calculations showed that the configuration with a tilted water
molecule at the $T_1$ site on Cu(100) would be more stable than that with
the axis of the water molecule parallel to the surface normal,%
\cite{Ribarsky,Ignaczak} they predicted that the $T_1$ site would not be
the energetically most favorable bonding site. Instead, the calculations  
showed that water would prefer to bond at the two-fold bridge ($B_2$) site
with its molecular axis parallel to the surface normal.\cite{Ignaczak}
Similarly, semi-empirical quantum chemical cluster calculations concluded
that the adsorption of a water molecule at a four-fold hollow site ($H_4$)
would be energetically most favorable.\cite{Kuznetsov} Both results
disagree with the experimental data, which suggest a tilted geometry of  
water at the $T_1$ site on Cu(100).

Here we present results of total-energy density-functional calculations in
which supercell models are used to simulate the Cu(100) surface. The
atomic and electronic structures of the adsorbed surface, as well as the 
preferred bonding site and orientation of water molecules, are determined.
In agreement with the experimental data, the most stable adsorption site 
for water molecules chemisorbed on the Cu(100) surface is determined to be
the on-top $T_1$ site, and a tilted geometry of water at the on-top site
with an angle of approximately $70^\circ$ away from the surface normal is
shown to be energetically favorable. Moreover, rotational diffusion of the
water axis around the surface normal is predicted to dominate over lateral
diffusion at low temperatures. In addition, the energetics of molecularly
adsorbed water on the Cu(100) surface suggests a tendency for the
formation of water clusters on the surface.

The remainder of this paper is organized as follows. In Sec.~II we
describe the computational method and the supercell models that we used.
In Sec.~III we present and discuss our theoretical results. Finally, in 
Sec.~IV, we summarize the main conclusions obtained from our calculations.

\section{Method}

On the (100) surface of an f.c.c. metal, there are three different
symmetric adsorption sites, known as four-fold hollow ($H_4$), two-fold
bridge ($B_2$), and one-fold on-top ($T_1$) sites.\cite{Wang3} We studied
the adsorption of water at each of the three bonding sites on Cu(100), and
we also investigated the orientation of the water molecule on the surface.
The Cu(100) surface was modeled by repeated slabs with seven copper layers
separated by a vacuum region equivalent to nine copper layers. Each metal
layer in the supercell contained nine Cu atoms (a $3\times3$ surface unit
cell). Water was adsorbed symmetrically on both sides of the slab. All the
Cu atoms were initially located at their bulk positions, with the
equilibrium lattice constant of the bulk determined by our calculations.  
Calculations with a larger supercell containing nine copper layers showed
that convergence of the binding energy of a water molecule on the Cu(100)
surface for the most stable configuration with respect to the thickness of
the slab was to be within 0.02~eV.

The calculations were performed within density-functional theory, using
the pseudopotential method and a plane-wave basis set.%
\cite{Payne,Kresse1,Kresse2,Kresse3,Wang1,Wang2} The results reported in
this paper were obtained using the Vienna {\it ab-initio} simulation
package (VASP).\cite{Kresse1,Kresse2,Kresse3} The exchange-correlation  
effects were treated with the generalized gradient-corrected
exchange-correlation functionals (GGA) given by Perdew and
Wang.\cite{Perdew1,Perdew2} We adopted the Vanderbilt ultrasoft
pseudopotentials supplied by Kresse and Hafner.\cite{Vanderbilt,Kresse4}
A plane-wave energy cutoff of 30~Ry and 10 special {\bf k} points in the
irreducible part of the two-dimensional Brillouin zone of the $3\times3$
surface cell were used for calculating the H$_2$O-adsorbed Cu (100)
surface. Test calculations with cutoff energies up to 35~Ry and with 
different numbers of special {\bf k} points (6 and 10 points) obtained
the binding energy of a water molecule on the Cu(100) surface for the
most stable configuration within 0.01~eV. Optimization of the atomic 
structure was performed for each supercell {\it via} a conjugate-gradient
technique using the total energy and the Hellmann-Feynman forces on the
atoms.\cite{Payne} All the structures were fully relaxed until the forces
on all the atoms were less than 0.03~eV/{\AA}. In this sense, the O-H
bond length and the H-O-H angle of the H$_2$O molecule were optimized.  

\section{Results and discussions}

We first present the calculated properties for bulk copper and the relaxed
clean Cu(100) surface. Calculations for bulk Cu were conducted with 408
special {\bf k} points and a cutoff energy of 30~Ry. The total energy
convergence with respect to the cutoff energy (25-40~Ry) and the number 
of the {\bf k} points (220-570) was within a few tenths of 1~meV per atom.
We obtained lattice constants of 3.64~{\AA} for bulk Cu, in good agreement
with the experimental value of 3.61~{\AA}.\cite{Wyckoff}

\begin{table}[t]
\caption{Relaxation of the clean Cu(100) surface. $\Delta d_{ij}$ is the
change of the interlayer distance, and $d_{0}$ is the corresponding
distance in the bulk.}
\begin{ruledtabular}
\begin{tabular}{lcccccccc}
&${\Delta d_{12}}/d_0$ (\%)&&&${\Delta d_{23}}/d_0$ (\%)&&&${\Delta 
d_{34}}/d_0$ (\%)\\
\colrule
This work&$-$3.4&&&$+$1.5&&&$+$0.2\\
Experiment\footnotemark[1]&$-$2.8&&&$$+$1.1$&&&$\pm$0.0\\  
\end{tabular}
\end{ruledtabular}
\footnotetext[1]{Ref. \onlinecite{Jiang}.}
\end{table}

\begin{figure}[b]
\includegraphics{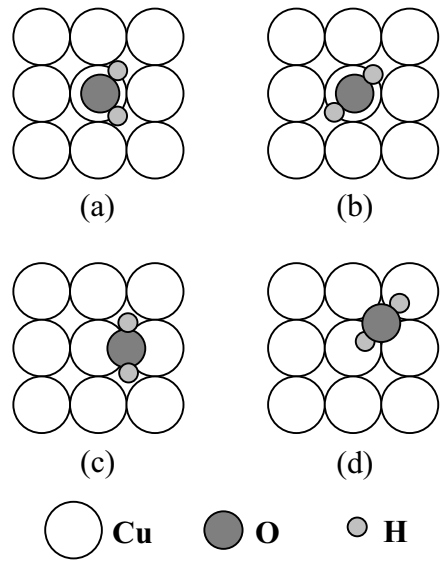}
\caption{Schematics of the minimum-energy structures (top view)
corresponding to the configurations with a water molecule at (a) the $T_1$
site (with a tilted geometry), (b) the $T_1$ site (without a tilted
geometry), (c) the $B_2$ site (without a tilted geometry), and (d) the
$H_4$ site (without a tilted geometry) on the Cu(100) surface.}
\label{autonum}
\end{figure}

The properties of the clean Cu(100) surface were calculated using a
$1\times1$ surface cell and 66 special {\bf k} points in the surface
Brillouin zone. All the seven copper layers except for the central one
were relaxed. The surface energy of the Cu(100) surface was calculated to
be 0.59 eV, and the results for the surface relaxation are shown in Table
1. The surface shows an inward relaxation (approximately $0.03$~{\AA}) of
the top layer and an outward relaxation (also approximately $0.03$~{\AA})
of the second layer. Our results, in good agreement with LEED 
measurements,\cite{Jiang} show relaxation of the Cu(100) surface with
${\Delta d_{12}}/d_0 = -3.4 \%$ (contraction) and  ${\Delta d_{23}}/d_0 =
+1.5 \%$ (expansion), where ${\Delta d_{12}}$ and ${\Delta d_{23}}$ are 
the changes in spacing between the top and the second layer and between 
the second and the third layer, and $d_0$ is the bulk interlayer distance.

Table II shows the calculated binding energies of a water molecule in the
different configurations of the H$_2$O-adsorbed Cu(100) surface. The most
stable configuration, shown in Fig.~1(a), is that with a tilted water
molecule at the on-top $T_1$ site (tilted-$T_1$ configuration). The water
molecule is bonded to the surface with the oxygen end toward a surface Cu 
atom and the molecular axis tilted away from the surface normal at an
angle of 73$^\circ$. The binding energy of a water molecule in this
configuration is calculated to be 0.25~eV. Three metastable configurations
(local energy minima) were determined from our calculations: a water
molecule with its axis parallel to the surface normal at the $T_1$
(Fig.~1(b)), $B_2$ (Fig.~1(c)), and $H_4$ (Fig.~1(d)) sites, respectively.
The binding energies of a water molecule in these metastable
configurations are smaller by 0.08-0.21~eV than that in the tilted-$T_1$
configuration. We also optimized the structures of the tilted-$B_2$ and
tilted-$H_4$ configurations (a tilted water molecule at the $B_2$ and
$H_4$ sites, respectively), and found that neither was a local energy
minimum. In fact, the tilted water molecule in such configurations was
found to relax eventually to a nearby $T_1$ site with a tilted geometry,
that is, the tilted-$T_1$ configuration (Fig.~1(a)). We therefore conclude
that water adsorbed on the Cu(100) surface prefers the $T_1$ site with its
axis tilted away from the surface normal. This conclusion is in agreement
with the experimental observations.

\begin{table}[t]
\caption{The binding energies (in unit of eV) of the water molecule
adsorbed in the different configurations on the Cu(100) surface. $\psi$ is
the angle between the surface normal and the axis of the molecular water.}
\begin{ruledtabular}
\begin{tabular}{ccccccccc}
$T_1~(\psi = 73^\circ)$&$T_1~(\psi = 0^\circ)$&$B_2~(\psi =
0^\circ)$&$H_4~%
(\psi = 0^\circ)$\\
\colrule
0.25&0.17&0.14&0.04\\
\end{tabular}
\end{ruledtabular}
\end{table}

\begin{figure}[b]
\includegraphics{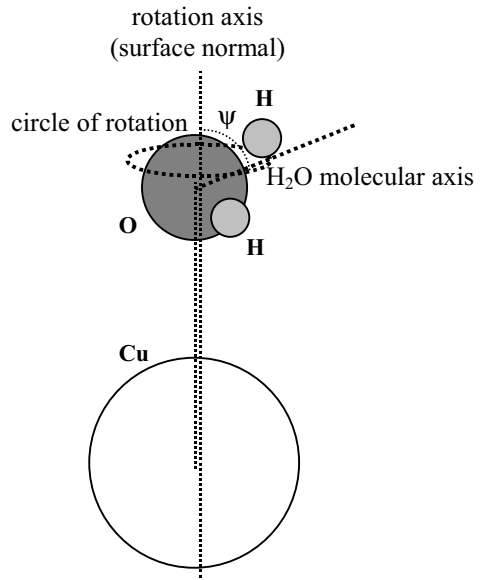}
\caption{Schematics of the rotation of the water molecule around the
surface normal in the tilted $T_1$ configuration on the Cu(100) surface.
Note that the O-Cu connecting line is slightly off (by approximately
0.03~{\AA}) the rotation axis and rotates around it.}
\label{autonum}
\end{figure}

Previous theoretical studies employing cluster models also determined the
preferred bonding sites of water on the Cu(100) surface, as mentioned in
Sec.~I. {\it Ab initio} DFT calculations using a 13-atom cluster for the
Cu(100) surface showed that water would prefer to bond at the $B_2$ site 
with its molecular axis normal to the surface.\cite{Ignaczak}
Semi-empirical quantum chemical cluster calculations with a 9-atom cluster
predicted that the binding energy of a water molecule at the $H_4$ site  
was significantly larger than at the $T_1$ site (by approximately 
$0.7$~eV).\cite{Kuznetsov} These theoretical conclusions are inconsistent
with both our {\it ab initio} supercell calculations and the experimental
data. The main problem is that the previous calculations were limited to
small clusters, containing only up to 13 metal atoms. In contrast to
semiconductors, metals are often poorly represented by small clusters.%
\cite{Whitten,Panas,Whetten,Gensic,Wang4a,Wang4b,Mitchell,Wang4,Wang5,%
Dominguez} Large clusters or extended surface models are usually needed  
to simulate real metal surfaces in order to achieve good convergence in
calculations.

The obtained structural parameters of the optimized geometry for the 
tilted-$T_1$ configuration show that the water molecule keeps essentially
the same shape upon adsorption on the surface with the equilibrium O-H
bond length (0.98~{\AA}) and H-O-H angle (105.6$^\circ$) slightly larger
than those of a free molecule: 0.97~{\AA} and 105.3$^\circ$, respectively,
determined from our calculations.\cite{Note1} The experimental
values\cite{Eisenberg} of the O-H bond length and H-O-H bond angle for an
isolated water molecule are 0.96~{\AA} and 104.5$^\circ$, respectively.
After the adsorption of a water molecule, the surface Cu atom that is
bonded to oxygen is observed to shift its position up along the surface  
normal (by 0.08~{\AA}) and also to have a slight in-plane shift of
0.02~{\AA} along the direction against the H$_2$O molecular axis. The   
distance between the oxygen and its bonded surface copper atom is
2.28~{\AA}. The structural changes reflect the fact that there is a
considerable (though weak) interaction between the adsorbed water molecule
and the Cu(100) surface.

It is interesting to note that the variation of the total energies for the
tilted-$T_1$ configurations with the rotation of the water molecule around
the surface normal (shown in Fig.~2) was determined to be less than 
0.02~eV. The activation energy for rotation is thus about 0.08~eV smaller
than the lateral diffusion barrier for a water molecule on the Cu(100) 
surface, which was estimated to be approximately 0.1~eV (with
$T_1$-$B_2$-$T_1$ taken as the diffusion pathway). While there so far have
been no experimental efforts to observe such rotational diffusion, these
energetic considerations suggest that it should occur at much lower
temperatures than lateral diffusion. Lateral diffusion at elevated   
temperatures has been experimentally observed for single water molecules 
initially adsorbed at low temperatures of about 20~K.\cite{Anderson} The
occurrence of the rotation of the water molecule around a stiff axis on 
the surface is a consequence of the fact that the bonding of water with   
the Cu(100) surface is dominated by the oxygen lone-pair electronic
orbital (see below).

Reliable theoretical values for the water sublimation energy have been
obtained with {\it ab initio} DFT calculations.\cite{Hamann,Feibelman}
Hamann obtained a value of 0.66~eV for the H$_2$O ice sublimation energy,%
\cite{Hamann} while Feibelman,\cite{Feibelman} using the same computation
package (VASP) as we employed, obtained values of 0.67-0.72~eV for the
D$_2$O ice sublimation energy.  Our calculated binding energy (0.25~eV)
of H$_2$O on Cu(100) is thus significantly smaller than the water
sublimation energy. Therefore, water molecules apparently tend to form
ice-like clusters on the Cu(100) surface. This could occur even at low
water coverages when high temperatures make the water molecules diffuse on
the surface. Experimental measurements have demonstrated that water
clusters were indeed formed on the Cu(100) surface.%
\cite{Anderson,Spitzer,Nyberg,Brosseau} In particular, only one peak was
observed at a temprature of approximately 160~K in the TPD spectrum of
H$_2$O on Cu(100), indistinguishable from the ice sublimation peak.%
\cite{Sexton,Brosseau}

\begin{figure}[t]
\includegraphics{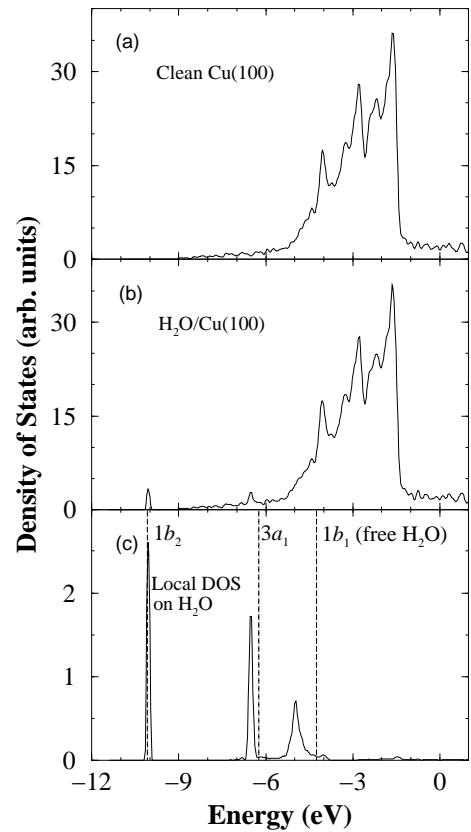}
\caption{Total density of states (DOS) for the clean (a) and the
water-adsorbed (b) Cu(100) surfaces and local DOS on the water molecule
(c) for the water-adsorbed Cu(100) surface. The Fermi level is at 0 eV.
The calculated energy levels for the molecular orbitals of a free water  
molecule are also indicated in (c).}
\label{autonum}
\end{figure}

The calculated electronic density of states (DOS) for the clean and the
H$_2$O-adsorbed Cu(100) surfaces are shown in Fig.~3. The main changes of
the DOS upon the adsorption of water are found to be in the energy region
lower than about 4~eV below the Fermi level on the DOS curve. Two new    
peaks (Fig.~3(b)), located at $-$10.07 and $-$6.52~eV relative to the
Fermi level, respectively, are clearly observed. The curve for the
projected density of states (projected or local DOS) onto the water   
molecule (Fig.~3(c)) indicates that these two peaks correspond to the  
water 1$b_2$ and 3$a_1$ states. The highest-energy electronic state of
water, that is, the 1$b_1$ molecular orbital, is buried in the substrate
energy band and hardly observed in the total DOS curve. However, the local
DOS curve shows that its peak is located at $-$4.97~eV relative to the
Fermi level (Fig.~3(c)). Upon adsorption, the 1$b_1$ and 3$a_1$ orbitals
of the water molecule are observed to shift down in energy by 0.72~eV and
0.29~eV, respectively, while the 1$b_2$ orbital remains almost unchanged
in energy level. The 1$b_1$ state is also seen to be more delocalized when
water is adsorbed on the surface. These results suggest that water is
bonded to the surface {\it via} the 1$b_1$ state (with some contributions
from the 3$a_1$ state). The 1$b_1$ state is primarily of oxygen 2$p$
lone-pair character, and a small charge transfer from water to the copper
surface is thus mostly from the oxygen 2$p$ state (water is known as a 
Lewis base). Such a charge transfer and the resulting bonding between the
1$b_1$ state and the surface would be facilitated when the 2$p$ orbital is
perpendicular to the surface and toward a surface atom. Therefore, a water
molecule would prefer to bond {\it via} oxygen at the on-top $T_1$ site on
the Cu(100) surface. Moreover, the molecular axis of water would be tilted
away from the surface normal since the 1$b_1$ orbital and the O-H bonding
orbitals in water tend to form a quasi-tetrahedral geometry. Thus, the   
tilted-$T_1$ configuration represents the most stable structure for the  
water-adsorbed Cu(100) surface. In addition, rotation of the water axis  
around the surface normal has essentially no effects on the bonding  
between the 1$b_1$ orbital and the surface, resulting in an very small
energy cost for such a rotation. These qualitative conclusions, based 
on the DOS and electronic structures, are in excellent agreement with the
quantitative results of the energetics discussed above.

\section {Conclusions}

The theoretical approach of supercell models combined with
first-principles total-energy DFT pseudopotential methods has reproduced
experimental measurements of the preferred adsorption geometry for the
water-adsorbed Cu(100) surface. Very small differences in total energies
among configurations with different orientations and bonding sites of the
water molecule have been distinguished. It is shown that the on-top $T_1$
configuration is the most stable structure for the H$_2$O-adsorbed Cu(100)
surface, and the water molecule is shown to be tilted with its axis
approximately $70^\circ$ away from the surface normal. The calculations
also predict that rotational diffusion of the water axis around the surface
normal would occur at low temperatures. In addition, water molecules are
likely to form ice-like clusters on the Cu(100) surface at high coverages
or at high temperatures.

\acknowledgments

This work was supported by NSF Grants DMR-0111841 and DMR-0240078 and by   
the School of Computational Science and Technology (CSIT) at Florida State
University. Access to the CSIT SP3 and SP4 supercomputers is also
acknowledged.

\end{document}